# Chiral Multi-layer Waveguides Incorporating Graphene Sheets: An Analytical Approach

Mohammad Bagher Heydari, Mohammad Hashem Vadjed Samiei

*Abstract*—**This paper presents a novel analytical model for chiral multi-layer waveguides incorporating graphene sheets. The general structure is composed of various chiral layers, where a graphene sheet has been sandwiched between two adjacent chiral layers. We have chosen and studied this general structure to suggest a new platform for novel emerging sciences such as optical sensing. An analytical model has been proposed for the general structure to find its dispersion relation and other modal properties such as the effective index and the propagation length. To show the richness of our general structure, two novel chiral waveguides containing graphene sheets have been introduced and investigated. Our analytical results show that high values of effective indices, e.g. $n_{eff}$ = 15 at the frequency of 8 THz for the second structure, is achievable. Moreover, it has been shown that the chemical potential of the graphene and the chirality can adjust and control the plasmonic features. Our multi-layer structure and its analytical model can be utilized in several potential applications such as switches, absorbers, cloaks, polarization rotators, and directional couplers in the THz frequencies.**

*Index Terms*—**Analytical model, Graphene layer, Chiral medium, Multi-layer structure**

## I. INTRODUCTION

IN RECENT YEARS, chiral materials have attracted the attention of many scientists due to their interesting features such as optical activity and circular dichroism [1-6]. Based on these fascinating properties, various chiral-based devices have been proposed and reported such as polarization rotators [7], cloaks [8], and planar waveguides [9-11]. The chiral medium supports hybrid modes, which are split into two propagating modes as the frequency increases. These modes are known as "bifurcated modes" in the literature and have many applications such as chiro-waveguides [12], parallel plate chiral structures [13, 14], and microstrip antennas [15].

Nowadays, a new branch of optics called "Plasmonics" is emerged [16] and many research articles have been reported in this field such as photo-detectors [17-21] and plasmonic waveguides [22, 23]. Graphene is one of the two-dimensional materials that can support tunable high-quality plasmons, i.e. high effective index and low propagation loss, in mid-infrared frequencies [24]. The important plasmonic feature of graphene is its optical conductivity, a tunable parameter via electrostatic (or chemical potential) and magnetostatic biasing, that allows one to propose novel THz components such as plasmonic waveguides [25-34], isolator [35], coupler [36], resonator [37], antennas [38-40], filter [41], circulators [42, 43], Radar Cross-Section (RCS) reduction-based devices [44-46], and graphene-based medical components [47-53]. It should be noted that noble metals support SPPs at the near-infrared and visible frequencies [37, 54, 55]. However, some significant features of graphene-based components, such as extreme confinement, tunable conductivity, and low losses in THz and mid-infrared frequencies differ from any metal-air interface waveguides [56-65].

The hybridization of graphene and chiral medium can be an interesting matter because it gives more degrees of freedom to the designer for effectively controlling and tuning the modal properties of the structure by changing the chemical potential of graphene and the chirality parameter of the chiral medium. Some plasmonic structures based on chiral-graphene hybridization have been addressed in several articles [26, 66, 67]. In [66], the authors have presented a dispersion relation for a graphene-chiral interface and then have studied the propagation properties of the interface. A graphene-chiral-graphene slab waveguide is reported in [26], which only investigates the variations of the determinant of the dispersion matrix as a function of the propagation constant and not enough study is done on the plasmonic features of the graphene-chiral slab in this work. The authors in [67] have introduced a chiral-graphene-metal structure for the optical frequency range and have considered the influence of various parameters on the dispersion diagram.

To cover all special cases of the graphene-based chiral structures, we present a novel analytical model for general chiral multi-layer structures containing graphene sheets. To the best of our knowledge, a comprehensive study on the plasmonic features of chiral multi-layer waveguides incorporating graphene layers has not been investigated in any published paper. Our general structure is composed of various chiral layers, each chiral layer has the permittivity, the permeability and the chirality of $\varepsilon_c, \mu_c, \gamma_c$, respectively. A graphene sheet,

Mohammad Bagher Heydari and Mohammad Hashem Vadjed Samiei are with the School of Electrical Engineering, Iran University of Science and Technology (IUST), Tehran, Iran (e-mail: mo_heydari@elec.iust.ac.ir).

with an isotropic conductivity of $\sigma$, has been sandwiched between two adjacent chiral layers. The proposed rich structure gives the designer more degrees of freedom to effectively control and change the modal properties by altering the chemical potentials of the graphene sheets ($\mu_{g,1}$, $\mu_{g,2}$, $\mu_{g,3}$, ...) and the chirality of chiral layers ($\gamma_{c,1}$, $\gamma_{c,2}$, $\gamma_{c,3}$, ...). The structure supports hybrid plasmonic modes due to the existence of chiral layers. These modes are split into two propagating modes, which we will call them "Higher modes" and "Lower modes" in this article. Higher modes and lower modes are related to the high and low plasmon resonance frequencies, respectively. It is worthwhile to mention that the chirality has a vital role in life sciences so that many fascinating applications have been reported for chiral-based detection and sensing [68-71]. Besides, graphene is an easy-available material in the market and it has tunable conductivity, which can generate, control and modulate high-quality surface plasmon polaritons (SPPs). Therefore, a hybrid graphene-chiral structure can be a novel and promising platform for new emerging sciences such as optical sensing.

The paper is organized as follows. In section II, we will introduce the general chiral multi-layer structure containing graphene sheets and will propose a new analytical model for it. This section obtains a matrix representation for the general structure that its determinant gives the dispersion relation of the multi-layer waveguide. In section III, before embarking on numerical simulations of two special cases of chiral multi-layer waveguides, we will depict the effective index of TM plasmonic mode for a familiar graphene-based achiral waveguide, constituting Air-Graphene-SiO$_2$-Si layers, to show high accuracy of our analytical model. A full agreement is seen between the analytical results of our model and the numerical result of [72], which confirms the high accuracy of the analytical model outlined in section II. To show the richness of the general structure, two new cases of the graphene-chiral waveguide, as special cases of the general structure, are studied in section III. The first case is a chiral slab waveguide, where the graphene sheet has been deposited on chiral-SiO$_2$ layers. The second waveguide is a novel hybrid chiral-graphene structure, constituting graphene-chiral-graphene-chiral-SiO$_2$ layers, which applies two graphene sheets together with two chiral layers. Compared to the first structure, the analytical results indicate a high performance for the second waveguide and it has more degrees of freedom for tuning the modal propagating properties. To the authors' knowledge, our proposed graphene-chiral slab waveguides have not been presented in any published work. Finally, section IV concludes the article.

## II. THE PROPOSED GENERAL STRUCTURE AND ITS ANALYTICAL MODEL

Fig. 1 illustrates the schematic of the general structure, where each graphene sheet has been sandwiched between two adjacent chiral layers. The conductivity of the graphene sheet is modeled by Kubo's relation [73]:

$$\sigma_N\left(\omega, \mu_{g,N}, \Gamma_N, T\right) = \frac{-je^2}{4\pi\hbar} Ln\left[\frac{2\left|\mu_{g,N}\right| - (\omega - j2\Gamma_N)\hbar}{2\left|\mu_{g,N}\right| + (\omega - j2\Gamma_N)\hbar}\right] +$$
$$\frac{-je^2 K_B T}{\pi\hbar^2(\omega - j2\Gamma_N)}\left[\frac{\mu_{g,N}}{K_B T} + 2Ln\left(1 + e^{-\mu_{g,N}/K_B T}\right)\right] \tag{1}$$

In the above relation, $N$ indicates the index of the graphene layer (As seen in Fig. 1, each graphene sheet has different conductivity in the general). Furthermore, $\hbar$ is the reduced Planck's constant, $K_B$ is Boltzmann's constant, $\omega$ is radian frequency, $e$ is the electron charge, $\Gamma_N$ is the phenomenological electron scattering rate for the $N$-th layer ($\Gamma_N = 1/\tau_N$, where $\tau_N$ is the relaxation time), $T$ is the temperature, and $\mu_{g,N}$ is the chemical potential for the $N$-th layer which can be altered by chemical doping or electrostatic bias [73].

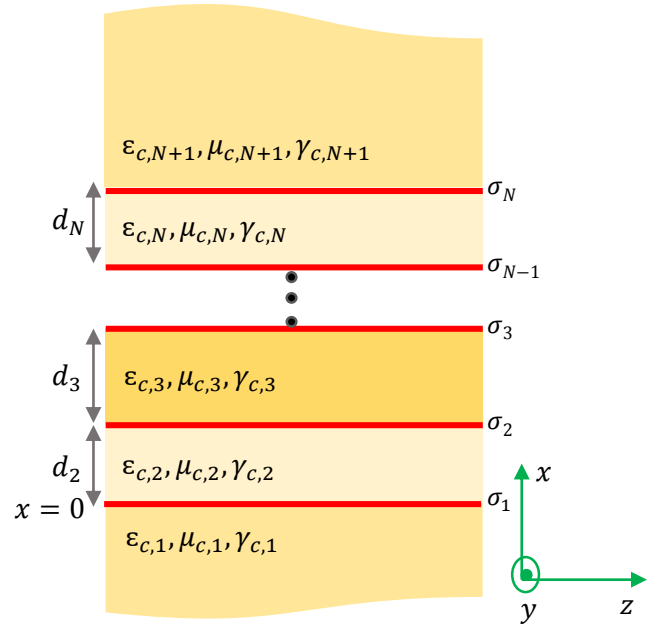

Fig. 1. The cross-section of the general proposed structure.

Let us assume that each chiral layer has the permittivity, the permeability and the chirality of $\varepsilon_{c,N}$, $\mu_{c,N}$, $\gamma_{c,N}$, respectively, which satisfy the following relations [12]:

$$\boldsymbol{D}_N = \varepsilon_{c,N}\boldsymbol{E}_N - j\gamma_{c,N}\boldsymbol{B}_N \tag{2}$$

$$\boldsymbol{H}_N = -j\gamma_{c,N}\boldsymbol{E}_N + \frac{\boldsymbol{B}_N}{\mu_{c,N}} \tag{3}$$

In the above relations, $\varepsilon_0$, $\mu_0$ are the permittivity and permeability of free space, respectively.

As obtained in [74], the electromagnetic fields inside the chiral medium are the superposition of left-handed circularly polarized (LCP) and right-handed circularly polarized (RCP) waves. We suppose that the electromagnetic fields propagate in the z-direction ($e^{j\beta z - i\omega t}$) and also they have no variations in the y-direction ($\frac{\partial}{\partial y} = 0$). Therefore, the Helmholtz's equation for the $N$-th layer is written as:

$$\frac{d^2\Psi_{R,N}}{dx^2} + q_{R,N}^2 \Psi_{R,N} = 0 \tag{4}$$

$$\frac{d^2\Psi_{L,N}}{dx^2} + q_{L,N}^2 \Psi_{L,N} = 0 \tag{5}$$

where

$$q_{R,N} = \sqrt{\beta^2 - k_{R,N}^2} \tag{6}$$

$$q_{L,N} = \sqrt{\beta^2 - k_{L,N}^2} \tag{7}$$

In (6)-(7), the propagation constants of RCP and LCP waves have been defined as follows:

$$k_{R,N} = \omega\left(n_{c,N} + \gamma_{c,N}\right) \tag{8}$$

$$k_{L,N} = \omega\left(n_{c,N} - \gamma_{c,N}\right) \tag{9}$$

Where $n_{c,N}$ is the refractive index of the $N$-th layer of chiral medium:

$$n_{c,N} = \sqrt{\frac{\mu_{c,N}\,\varepsilon_{c,N}}{\mu_0\,\varepsilon_0}} \tag{10}$$

Then, the Helmholtz's functions are written by adding LCP and RCP wave-functions:

$$\Psi_{R,N} = A_{R,N}\cos\left(q_{R,N}x\right) + B_{R,N}\sin\left(q_{R,N}x\right) \tag{11}$$

$$\Psi_{L,N} = A_{L,N}\cos\left(q_{L,N}x\right) + B_{L,N}\sin\left(q_{L,N}x\right) \tag{12}$$

In (11)-(12), $A_{R,N}, B_{R,N}, A_{L,N}, B_{L,N}$ are unknown coefficients and are obtained by applying boundary conditions. In the chiral medium, the electromagnetic fields can be written as:

$$\Psi_{R,N} = E_{z,N} + j\eta_{c,N}H_{z,N} \tag{13}$$

$$\Psi_{L,N} = E_{z,N} - j\eta_{c,N}H_{z,N} \tag{14}$$

Which yields to

$$E_{z,N} = \frac{\Psi_{R,N} + \Psi_{L,N}}{2} \tag{15}$$

$$H_{z,N} = \frac{\Psi_{R,N} - \Psi_{L,N}}{2j\eta_{c,N}} \tag{16}$$

In the above relations, the impedance of the $N$-th chiral layer has been defined:

$$\eta_{c,N} = \sqrt{\frac{\mu_{c,N}}{\varepsilon_{c,N}}} \tag{17}$$

The transverse components of electric and magnetic fields are expressed as:

$$E_{T,N} = \begin{pmatrix} E_{x,N} \\ E_{y,N} \end{pmatrix} = -\frac{1}{2}\begin{pmatrix} j\beta & j\beta \\ k_R & k_L \end{pmatrix} \cdot \frac{\partial}{\partial x}\begin{pmatrix} \dfrac{\Psi_{R,N}}{q_{R,N}^2} \\ \dfrac{\Psi_{L,N}}{q_{L,N}^2} \end{pmatrix} \tag{18}$$

$$H_{T,N} = \begin{pmatrix} H_{x,N} \\ H_{y,N} \end{pmatrix} = -\frac{1}{2\eta_{c,N}}\begin{pmatrix} \beta & -\beta \\ -jk_R & jk_L \end{pmatrix} \cdot \frac{\partial}{\partial x}\begin{pmatrix} \dfrac{\Psi_{R,N}}{q_{R,N}^2} \\ \dfrac{\Psi_{L,N}}{q_{L,N}^2} \end{pmatrix} \tag{19}$$

By defining

$$\overline{\overline{Q}}_N^E = \begin{pmatrix} j\beta & j\beta \\ k_R & k_L \end{pmatrix} \tag{20}$$

$$\overline{\overline{Q}}_N^H = \begin{pmatrix} \beta & -\beta \\ -jk_R & jk_L \end{pmatrix} \tag{21}$$

$$\Phi_N = \begin{pmatrix} \dfrac{\Psi_{R,N}}{q_{R,N}^2} \\ \dfrac{\Psi_{L,N}}{q_{L,N}^2} \end{pmatrix} \tag{22}$$

The above relations for transverse components are written as:

$$E_{T,N} = -\frac{1}{2}\overline{\overline{Q}}_N^E \cdot \frac{\partial \Phi_N}{\partial x} \tag{23}$$

$$H_{T,N} = -\frac{1}{2\eta_{c,N}}\overline{\overline{Q}}_N^H \cdot \frac{\partial \Phi_N}{\partial x} \tag{24}$$

Now, we suppose that the propagation constants of RCP and LCP waves are following constants for various regions:

$$q_R = \begin{cases} jq_{R,1} & \text{for first layer} \\ q_{R,2} & \text{for second layer} \\ ... \\ q_{R,N-1} & \text{for } (N-1)-th\ layer \\ q_{R,N} & \text{for } (N)-th\ layer \\ jq_{R,N+1} & \text{for } (N+1)-th\ layer \end{cases} \tag{25}$$

$$q_L = \begin{cases} jq_{L,1} & \text{for first layer} \\ q_{L,2} & \text{for second layer} \\ ... \\ q_{L,N-1} & \text{for } (N-1)-th\ layer \\ q_{L,N} & \text{for } (N)-th\ layer \\ jq_{L,N+1} & \text{for } (N+1)-th\ layer \end{cases} \tag{26}$$

Hence, the Helmholtz's relations for various regions are written as:

$$\Psi_R = \begin{cases} A_{R,1}\exp\left(q_{R,1}x\right) & x < 0 \\ A_{R,2}\cos\left(q_{R,2}x\right) + \\ B_{R,2}\sin\left(q_{R,2}x\right) & 0 < x < d_2 \\ ... \\ A_{R,N-1}\cos\left(q_{R,N-1}x\right) + \\ B_{R,N-1}\sin\left(q_{R,N-1}x\right) & \sum_{k=2}^{N-2}d_k < x < \sum_{k=2}^{N-1}d_k \\ A_{R,N}\cos\left(q_{R,N}x\right) + \\ B_{R,N}\sin\left(q_{R,N}x\right) & \sum_{k=2}^{N-1}d_k < x < \sum_{k=2}^{N}d_k \\ B_{R,N+1}\exp\left(-q_{R,N+1}x\right) & x > \sum_{k=2}^{N}d_k \end{cases} \tag{27}$$

$$
\Psi_L = \begin{cases}
A_{L,1}\exp\left(q_{L,1}x\right) & x < 0 \\
A_{L,2}\cos\left(q_{L,2}x\right)+ & \\
B_{L,2}\sin\left(q_{L,2}x\right) & 0 < x < d_2 \\
... & \\
A_{L,N-1}\cos\left(q_{L,N-1}x\right)+ & \\
B_{L,N-1}\sin\left(q_{L,N-1}x\right) & \sum_{k=2}^{N-2}d_k < x < \sum_{k=2}^{N-1}d_k \\
A_{L,N}\cos\left(q_{L,N}x\right)+ & \\
B_{L,N}\sin\left(q_{L,N}x\right) & \sum_{k=2}^{N-1}d_k < x < \sum_{k=2}^{N}d_k \\
B_{L,N+1}\exp\left(-q_{L,N+1}x\right) & x > \sum_{k=2}^{N}d_k
\end{cases} \quad (28)
$$

Now, the longitudinal and transverse components of electromagnetic fields are derived for all layers by using relations (15)-(16) and (23)-(24). To obtain the characteristic equation, we should apply boundary conditions. In general form, the boundary conditions for a graphene sheet sandwiched between two adjacent chiral layers are written:

$$E_{z,N} = E_{z,N+1} \quad , \quad E_{y,N} = E_{y,N+1} \quad (29)$$

$$H_{z,N+1} - H_{z,N} = \sigma_N E_{y,N} \quad , \quad H_{y,N+1} - H_{y,N} = -\sigma_N E_{z,N} \quad (30)$$

At last, the final matrix representation is obtained:

$$
\overline{\overline{S}}_{4N,4N} \cdot \begin{pmatrix} A_{R,1} \\ A_{L,1} \\ A_{R,2} \\ B_{R,2} \\ A_{L,2} \\ B_{L,2} \\ ... \\ A_{R,N} \\ B_{R,N} \\ A_{L,N} \\ B_{L,N} \\ B_{R,N+1} \\ B_{L,N+1} \end{pmatrix}_{4N,1} = \begin{pmatrix} 0 \\ 0 \\ 0 \\ 0 \\ 0 \\ 0 \\ ... \\ 0 \\ 0 \\ 0 \\ 0 \\ 0 \\ 0 \end{pmatrix}_{4N,1} \quad (31)
$$

In (31), the matrix $\overline{\overline{S}}$ is:

$$
\overline{\overline{S}} = \begin{pmatrix}
\dfrac{1}{2} & \dfrac{1}{2} & ... & 0 & 0 \\
\dfrac{-1}{2q_{R,1}} & \dfrac{-1}{2q_{R,1}} & ... & 0 & 0 \\
... & ... & ... & ... & ... \\
0 & 0 & ... & ... & ... \\
0 & 0 & ... & ... & \dfrac{e^{-q_{L,N+1}\cdot\sum_{k=2}^{N}d_k}}{2}\left(\sigma_N+\dfrac{j}{q_{L,N+1}\eta_{c,N+1}}\right)
\end{pmatrix}
$$
$$(32)$$

Now, our analytical model has been completed for the general structure. By setting $\det\left(\overline{\overline{S}}\right) = 0$, the characteristic equation and then the propagation constant will be achieved. Afterwards, obtaining plasmonic features such as the effective index ($n_{eff} = Re[\beta]/k_0$) and the propagation length ($L_{Prop} = 1/(2Im[\beta])$) is straightforward. In what follows, we will investigate two chiral waveguides containing graphene sheets.

## III. SPECIAL CASES OF THE PROPOSED STRUCTURE: RESULTS AND DISCUSSIONS

Before embarking on numerical simulation of two special cases of chiral multi-layer waveguides, we depict the effective index of TM plasmonic mode for a familiar graphene-based achiral waveguide, constituting Air-Graphene-SiO₂-Si layers, to show high accuracy of our analytical model. In this structure, the permittivities of SiO₂ and Si layers are 2.09 and 11.9, respectively and parameters of the graphene sheet are $\mu_g = 0.23\ eV$, $\tau = 0.5\ ps$, $T = 3\ K$. The thickness of the SiO₂ layer is supposed to be 5µm. As observed in Fig. 2, the effective index increases as the frequency increases. A full agreement is seen between the analytical results of our model and the numerical result of [72], which confirms the high accuracy of the analytical model outlined in the previous section.

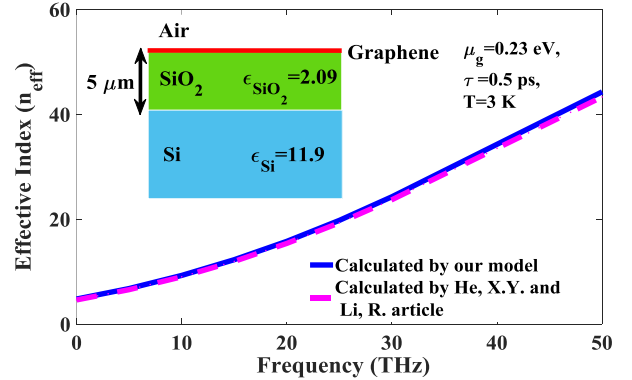

Fig. 2. The effective index of TM mode as a function of frequency for achiral waveguide constituting Air-SiO₂-Si layers, prepared by our analytical model and the results of [72]. The simulation parameters have been indicated in the diagram.

Now, we will introduce two novel graphene-based chiral slabs and will investigate their plasmonic features to show the richness of our proposed general structure. The first structure has graphene-chiral-SiO₂ layers, which support hybrid plasmonic modes. The second one has two chiral layers with different chirality parameters and thicknesses, constituting graphene-chiral-graphene-chiral-SiO$_2$ layers, which allows one to control and change the plasmonic properties such as the effective index by varying the chemical potentials of graphene sheets and the chirality of chiral layers.

In all following numerical simulations, the temperature is $T = 300\ K$ and the relaxation time of graphene layers is $\tau = 0.4\ ps$. Furthermore, the chemical potential of graphene sheets is assumed to be $\mu_g = 0.35\ eV$ unless otherwise stated.

It should be noted that hybrid modes in chiral-based waveguides are split into two propagating modes, which are known as "bifurcated modes" in the literature [75]. This phenomenon is one of the fascinating features of the chiral medium, which has been exploited in several applications such as chiro-waveguides [74], parallel plate chiral structures [13, 14], and microstrip antennas [15]. In this paper, we call these hybrid modes as "Higher modes" and "Lower modes", which are related to high and low plasmon resonance frequencies, respectively.

## A. The first structure: a chiral slab waveguide containing a graphene sheet

Fig. 3 illustrates the configuration of the first proposed waveguide, where a graphene sheet has been located on the chiral-SiO$_2$ layers. For simplicity and without the loss of generality, we suppose that the upper layer of the graphene is air. As already discussed in this paper, hybrid plasmonic modes, which are split into higher and lower modes, are excited in this structure due to the existence of the chiral slab. The chiral slab has a thickness of $d = 0.7\ \mu m$ and its effective index is $n_c = 1.3$. To show the tunability of plasmonic features by varying the chirality, various values have been considered for it ($\gamma = 0.002, 0.004, 0.006\ \Omega^{-1}$). The graphene parameters have been mentioned before.

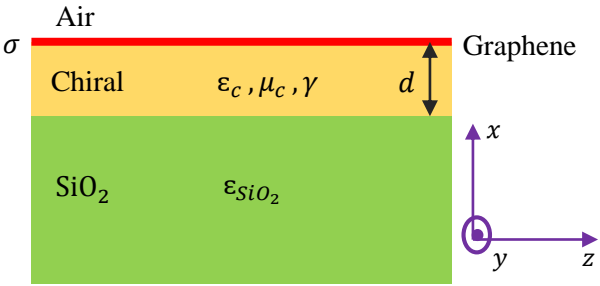

Fig. 3. The cross-section schematic of the first structure. The graphene sheet has been deposited on the chiral-SiO$_2$ layers.

Fig. 4 shows the analytical results of the effective index and the propagation length as a function of frequency for the first waveguide. As seen in Fig. 4(a), hybrid modes are split to higher and lower modes and the effective indices of them are tunable by changing the chirality. Fig. 4 clearly indicates the matter of why we called them "higher and lower modes". Consider a specific value of the propagation constant ($\beta$) or the effective index ($n_{eff}$), for instance $n_{eff} = 9$. For this propagation constant, the effective index diagram has two branches, where one branch appears at high frequencies (which we call them "higher modes") and the second one appears in low frequencies (which we call them "lower modes"). The effective index increases as the frequency increases, as seen in Fig. 4(a). One can observe from Fig. 4 (b) that the propagation length reduces with the frequency increment. Furthermore, for a specific frequency (for instance, consider $f = 3\ THz$), higher modes have high values of the propagation length compared to lower modes, but the effective index of lower modes is larger than higher modes.

Fig. 5 shows the normalized field distributions of $|E_x|, |E_y|$ and $|E_z|$. As seen in this figure, higher and lower modes have similar field distributions of $|E_z|$ and $|E_x|$. The only main difference between higher and lower modes is $|E_y|$, which can distinguish the different types of modes, i.e. higher and lower modes.

To see the effect of the chiral thickness on the effective index, the effective index has been depicted for the higher mode of the structure for various chiral thicknesses ($d = 0.7, 0.8, 0.9\ \mu m$) in Fig. 6. The chemical potential of the graphene sheet is 0.35 eV and the chirality of the chiral medium is assumed to be 0.002 $\Omega^{-1}$. As the thickness reduces, the effective index increases because the electromagnetic fields penetrate extremely inside the chiral layer for small chiral thicknesses.

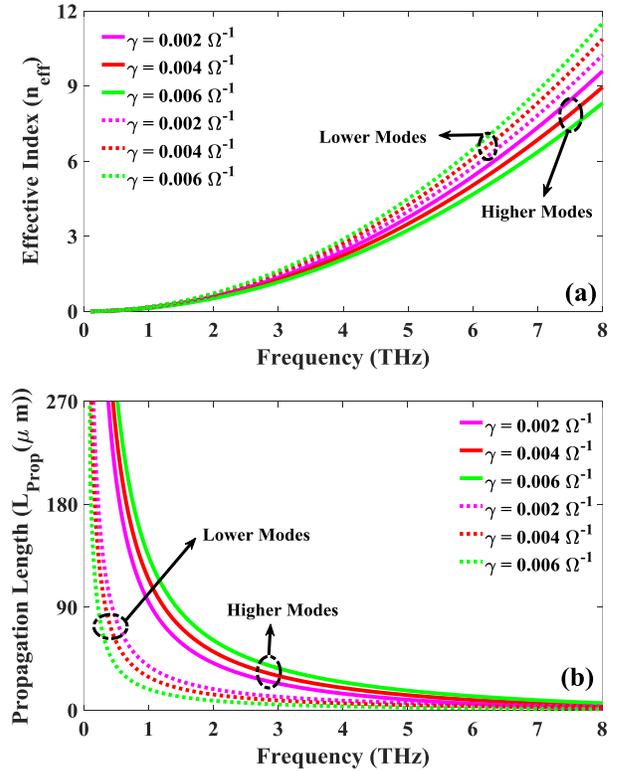

Fig. 4. Analytical results of plasmonic features for the first structure for various chirality values: (a) the effective index as a function of frequency for lower and higher hybrid modes, (b) the propagation length as a function of frequency for lower and higher hybrid modes. The chemical potential of graphene sheets is 0.35 eV.

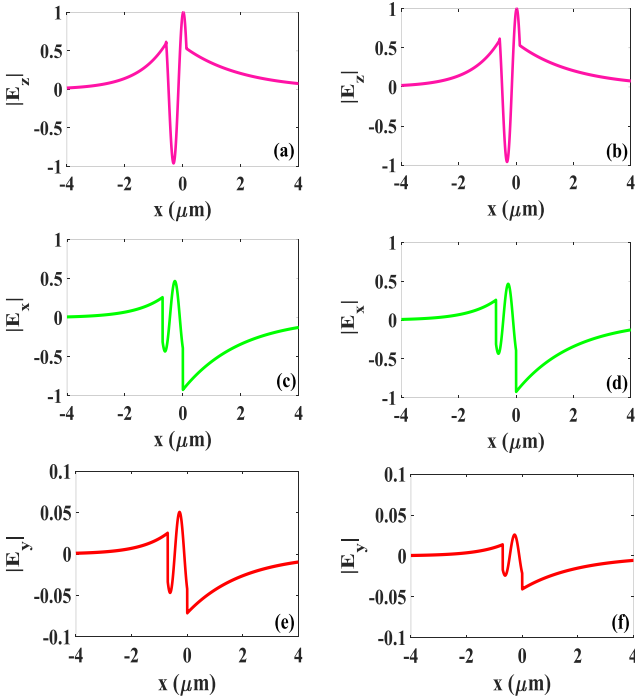

Fig. 5. Normalized field distributions of higher and lower modes: (a), (c), (e) higher mode, (b), (d), (f) lower mode, with $\mu_g = 0.35$ eV, $\gamma = 0.006$ $\Omega^{-1}$ and f = 6 THz.

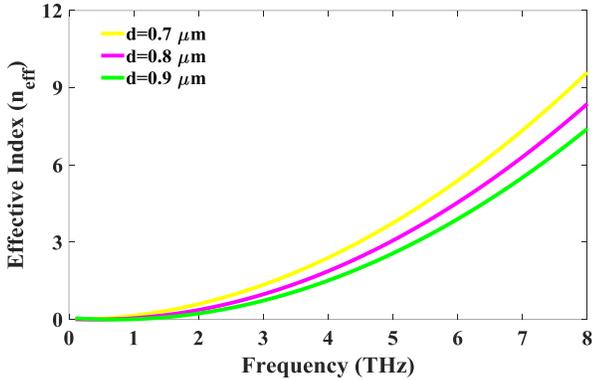

Fig. 6. The effective index of higher mode for the first structure for various chiral thicknesses ($d = 0.7, 0.8, 0.9 \mu m$). The chemical potential of the graphene sheet is 0.35 eV and the chirality of the chiral medium is 0.002 $\Omega^{-1}$.

## B. The second structure: hybrid graphene-based waveguide with two chiral layers

The schematic of the second structure, constituting graphene-chiral-graphene-chiral-SiO$_2$ layers, has been shown in Fig. 7. For simplicity, we assume that both graphene sheets have the same chemical potentials ($\mu_{g,1} = \mu_{g,2} = \mu_g$). The effective indices of first and second chiral slabs are $n_{c,1} = 1.3$, $n_{c,2} = 1.4$, respectively. In addition, the thicknesses of these layers are $d_1 = 0.7 \mu m, d_2 = 0.8 \mu m$. To show the tunability of plasmonic features by varying the chirality, various values have been considered for the chirality parameters of chiral layers ($\gamma_1 = 0.002, 0.004$ $\Omega^{-1}, \gamma_2 = 0.003, 0.005$ $\Omega^{-1}$).

Fig. 8 represents the plasmonic propagating properties as a function of frequency for the hybrid graphene-based chiral waveguide. Similar to the first structure, the hybrid plasmonic modes are transformed into higher and lower modes. It can be observed from Fig. 8(a) that the effective index increases as the frequency increases. Fig. 8(b) indicates that the propagation length reduces as the frequency increases. Furthermore, the propagation length of lower modes is not sensitive to the chirality parameters and their values are close together. A large value of the effective index, amounting to e.g. 15, is achievable at the frequency of 8 THz. Compared to the first structure, this waveguide has better performance (higher values of the effective index and propagation length). This high-performance originates from applying two graphene sheets and chiral slabs.

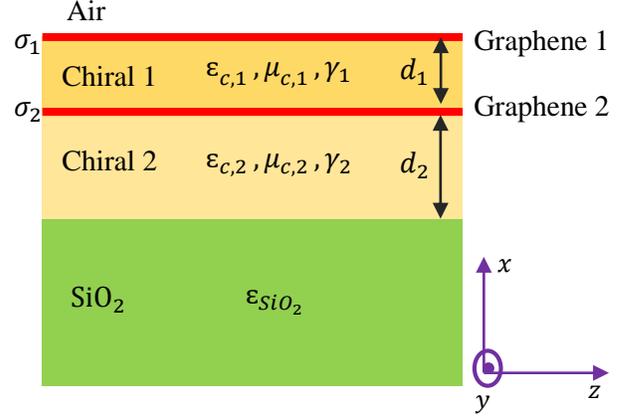

Fig. 7. The cross-section of the hybrid graphene-based waveguide with two chiral layers.

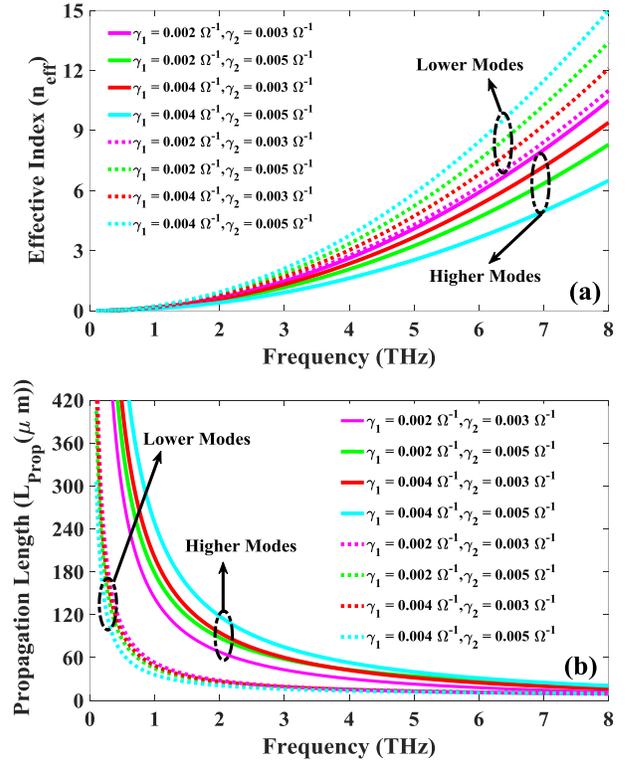

Fig. 8. Analytical results of plasmonic features for the second structure for various chirality values of two chiral layers: (a) the effective index as a function of frequency for lower and higher hybrid modes, (b) the propagation length as a function of frequency for lower and higher hybrid modes. The chemical potential of graphene sheets is 0.35 eV.

One of the important features of our proposed structure is its ability for changing the propagation properties by varying the chemical potential of the graphene sheets. To further study the influence of the chemical potential on the effective index, we have depicted the effective index as a function of the chemical potential for the various values of chirality at the frequency of 6 THz in Fig. 9. Again, it is supposed that both graphene sheets have the same chemical potentials ($\mu_{g,1} = \mu_{g,2} = \mu_g$) in this diagram. This figure gives important information to the designer for exploiting the structure at the desirable effective index. For instance, let us assume that we intend to design the waveguide for $n_{eff} = 9$. For this purpose, we can design the structure with one of these parameters: $\mu_g = 0.2\ eV$, $\gamma_1 = 0.002\ \Omega^{-1}$, $\gamma_2 = 0.003\ \Omega^{-1}$ or $\mu_g = 0.3 eV$, $\gamma_1 = 0.002\ \Omega^{-1}$, $\gamma_2 = 0.005\ \Omega^{-1}$ or $\mu_g = 0.23\ eV$, $\gamma_1 = 0.004\ \Omega^{-1}$, $\gamma_2 = 0.003\ \Omega^{-1}$ or $\mu_g = 0.41\ eV$, $\gamma_1 = 0.004\ \Omega^{-1}$, $\gamma_2 = 0.005\ \Omega^{-1}$.

Another fascinating application of Fig. 9 is determining the propagation range of hybrid modes for the specific chirality. For instance, the hybrid waveguide supports no modes for the chemical potential range of $\mu_g > 0.52\ eV$ at $\gamma_1 = 0.002\ \Omega^{-1}$, $\gamma_2 = 0.003\ \Omega^{-1}$. Furthermore, one can determine the effective indices of lower and higher modes for a specific chemical potential. Consider $\mu_g = 0.75\ eV$. For the chiral parameters of $\gamma_1 = 0.004\ \Omega^{-1}$, $\gamma_2 = 0.005\ \Omega^{-1}$, the effective indices of lower and higher modes are 2.9 and 3.2, respectively.

It should be mentioned that this figure represents a kind of Eigen-value problem. As the chemical potential increases, propagating modes lead to two modes "converging", which means that the behavior of two modes comes closer to each other for highly-doped graphene layers. At cross-over points, these modes degenerate into radiation modes in the cladding, and hence these plasmonic modes cannot be guided by the graphene layers.

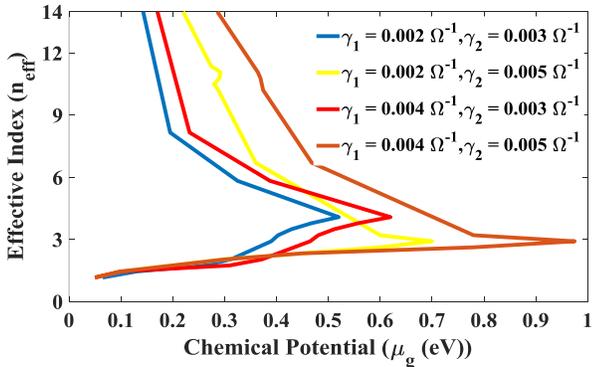

Fig. 9. The effective index of hybrid modes as a function of the chemical potential at the frequency 6 THz.

## IV. Conclusion

In this article, a new theoretical model is proposed for chiral multi-layer waveguides incorporating graphene sheets. The general structure supports hybrid bifurcated modes, which are split into higher modes and lower modes, with tunable modal features by changing the chemical potential and the chirality. Hybridization of chiral and graphene layers obtains a novel plasmonic platform, which can be utilized in new emerging sciences such as optical sensing. Two novel waveguides, as specific cases of the general structure, have been studied in this paper. The first structure is a chiral slab waveguide with a graphene sheet, where the graphene sheet has been placed on chiral-SiO$_2$ layers. The second one is a hybrid graphene-based chiral waveguide, constituting graphene-chiral-graphene-chiral-SiO$_2$ layers. The analytical results indicate that applying two graphene sheets together with two different chiral layers results in high performance for the waveguide. For instance, the effective index of 15 is reported for the second structure at the frequency of 8 THz. To show the tunability of the second structure, the effective index of plasmonic mode has been plotted as a function of chemical potential. This diagram could obtain important information to designers such as the propagation range of hybrid plasmonic mode for the specific chirality. For instance, no hybrid modes propagate for $\mu_g > 0.52\ eV$ at $\gamma_1 = 0.002\ \Omega^{-1}$, $\gamma_2 = 0.003\ \Omega^{-1}$ for the second structure at the frequency of 6 THz. In chiral-based multi-layer structures, the chirality parameters of each chiral medium, $\gamma_{c,1}, \gamma_{c,2}, \gamma_{c,3}, ...,$ give more degrees of freedom to the designer. Our analytical study for graphene-based chiral structures is very helpful for designing innovative THz devices such as directional couplers, switches, and absorbers.